# Première étape vers une navigation référentielle par l'image pour l'assistance à la conception des ambiances lumineuses

**Salma Chaabouni**[*] — **Jean-Claude Bignon**[*] — **Gilles Halin**[*]

[*] *CRAI ( Centre de Recherche en Architecture et Ingénierie) UMR M.A.P CNRS N°694*
*Ecole Nationale Supérieure d'Architecture de Nancy*
*2, rue Bastien Lepage, 54001 Nancy. France*

*chaabouni@crai.archi.fr, bignon@crai.archi.fr, halin@crai.archi.fr*

RÉSUMÉ. *Durant les premières phases de conception, l'image référence joue le double rôle de moyen de formulation et de résolution de problèmes. Dans notre approche, nous considérons l'image référence comme un outil de génération des idées et nous proposons un outil de navigation dans une base de références imagées pour assister la conception des ambiances lumineuses. Le présent article présente dans sa première partie la méthode d'indexation sémantique que nous avons employé pour l'indexation de notre base d'images. Il propose en deuxième partie une analyse synthétique de différents modes de navigation référentielle par l'image afin de proposer un outil d'assistance mettant en œuvre tout ou une partie de ces modes de navigation.*

ABSTRACT. *In the first design stage, image reference plays a double role of means of formulation and resolution of problems. In our approach, we consider image reference as a support of creation activity to generate ideas and we propose a tool for navigation in references by image in order to assist daylight ambience design. Within this paper, we present, in a first part, the semantic indexation method to be used for the indexation of our image database. In a second part we propose a synthetic analysis of various modes of referential navigation in order to propose a tool implementing all or a part of these modes.*

MOTS-CLÉS : *conception d'ambiance, image référence, indexation sémantique, navigation.*
KEY WORDS: *ambience design, image reference, semantic indexation, navigation .*

**1. Introduction**

Les simulations des phénomènes physiques de la lumière constituent une aide utile au contrôle de l'éclairage, mais sont peu appropriées à la recherche d'ambiances lumineuses aux premiers stades des projets d'architecture. Les outils de simulation de la lumière sont d'abord conçus comme des outils de vérification des solutions conçues [GRO 02]. Ils utilisent des paramètres d'évaluation qui supposent un état avancé de la conception. Par ailleurs, ces outils permettent essentiellement de faire des calculs de quantité de lumière dans un espace, ou de mettre en évidence la relation entre une source de lumière et une configuration spatiale mais ne prennent pas en compte la lumière en tant que matériau d'architecture pouvant participer à une mise en valeur de l'espace (valeur sensible de la lumière).

Dans cette recherche, nous considérons que pour instrumenter les premières étapes de la création, le concepteur fait moins appel à des outils de contrôle qu'à des heuristiques qui lui permettent d'avancer dans la formalisation de ses idées de conception. Parmi ces heuristiques, les raisonnements à base de références imagées [GUI 92] sont considérés comme particulièrement pertinents. La valeur cognitive de l'image référence dans le processus de conception constitue aujourd'hui un présupposé commun à des actions de recherches pourtant plurielles [GOL 95] [HAL 05], notamment dans le domaine des ambiances où plusieurs travaux se sont déjà intéressés à cette question [LAS 98] (Projet « Audience »[1]) (Projet « Daylight Design Variations Book »[2]). La valeur de projetation de l'image référence repose sur son double rôle de moyen de formulation de problème (les questions d'ambiance abordée par le concepteur) et de résolution de problèmes (les propositions faites par le concepteur). Elle fonctionne comme un modèle de questions et un modèle de solutions.

Dans notre approche, nous considérons l'image référence comme un outil de génération des idées de conception fondé sur des actions de comparaison, d'interprétation, de représentation, d'analogie… et nous proposons un outil de navigation dans une base de références imagées pour assister la conception des ambiances lumineuses.

Trois aspects indissociables coexistent dans notre travail, l'indexation, l'organisation et l'accès à l'information (images références).

Dans le présent article, nous nous intéresserons plus particulièrement à la méthode d'indexation que nous avons choisie puis nous présenterons une première approche sur des modes de recherche et de visualisation d'images références.

**2. Indexation d'images**

**2.1.** *Etat de l'art*

---

[1] http://audience.cerma.archi.fr/index.html
[2] http://sts.bwk.tue.nl/daylight/varbook/index.htm

L'indexation est souvent utilisée comme un ensemble de caractéristiques identifiantes dans une image. Elle résulte de l'analyse du contenu des images à indexer [DAU 94]. Il ne s'agit pas de coder toute l'information portée par l'image, mais de se concentrer sur l'information qui permet de traduire efficacement une réponse proche des besoins exprimés par un utilisateur. On distingue l'indexation par le contenu graphique et l'indexation par le contenu sémantique des images [KOU 07].

Dans la première catégorie, l'indexation est automatique et basée sur les caractéristiques visuelles de l'image à savoir la couleur, la texture et la forme. Plusieurs systèmes de recherche d'images par le contenu graphique ont été développés [VEL 01]. Ces systèmes permettent d'avoir des résultats pertinents lorsqu'une collection utilisée contient des images bien spécifiques comme les images de visages ou lorsqu'il s'agit d'une collection d'images très large comme les images de paysages. Cependant, ce type d'indexation fournit des résultats peu satisfaisants lorsqu'il s'agit de collection d'images hétérogènes [KAC 02].

Dans la deuxième catégorie -indexation sémantique- la description des images se fait avec des termes (mots clés) à partir d'une ontologie spécifique au domaine dont traitent les images [AMA 01]. Cette ontologie résulte de l'analyse du contenu des images.

**2.2. *Proposition***

Dans le cadre de notre approche, nous avons choisi une indexation sémantique afin de dépasser les difficultés de la recherche d'information dans des bases d'images hétérogènes. Nous avons commencé par une analyse des paramètres d'appréciation et de caractérisation des ambiances lumineuses pour la définition d'un vocabulaire d'indexation qui est organisé dans un thésaurus selon sept catégories.

Pour ce faire, nous avons analysé le vocabulaire employé par plusieurs architectes [CIR 91] et par les textes d'architecture en abordant la question des ambiances lumineuses. Le but a été de déterminer un ensemble de critères communs permettant de décrire différentes ambiances lumineuses représentées sur les images.

Au sein de notre laboratoire de recherche, nous avons développé un outil « iMage » qui nous permet l'indexation des images [Figure 1.]. Pour chaque image, nous avons distingué entre deux données résultantes de l'indexation.
  -les données informationnelles (non-visualisables)
  -les données visualisables de recherche (index)

Afin d'améliorer la finesse de l'indexation et pour tenir compte du fait qu'une image peut illustrer plus fortement certains concepts que d'autres nous avons ajouté une pondération des termes indexant l'image. A chaque terme est associé un poids marquant l'importance de ce terme dans la caractérisation de l'ambiance représentée sur l'image [VEN 06]. (Dans l'interface « iMage » la pondération se fait de une à quatre étoiles) .

Une représentation vectorielle des images est effectuée sur la base de cette indexation pondérée [SAL 83]. La similarité entre deux images est calculée à l'aide du modèle vectoriel.

Lors d'une requête, la mise en correspondance (calcul de similarité entre chaque image et la requête), donne pour chaque image un coefficient de similarité, qui permet de classer les images selon la pertinence qu'elles présentent pour l'utilisateur. Cette méthode permet également de naviguer sémantiquement dans une base de références imagées.

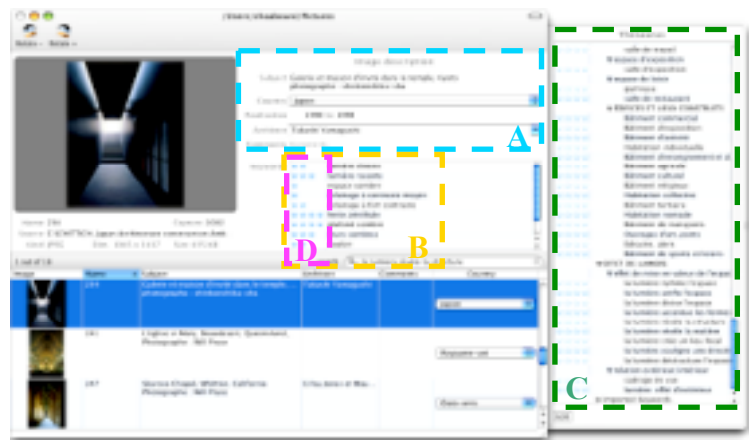

**Figure 1.** *Saisie d'écran de l'interface « iMage » représentant les données informationnelles « A », les index « B », le thésaurus « C » et la pondération « D »*

### 3. Vers une navigation sémantique et référentielle par l'image

Durant les phases amont de la conception, l'architecte-utilisateur peut rencontrer quelques difficultés à exprimer son besoin par des mots. Une interface de navigation référentielle par l'image doit permettre au concepteur d'exprimer son besoin en manipulant majoritairement des images et peu de texte.

Les premiers outils de manipulation de références réalisés dans notre laboratoire, Image.Idée et BatiMage, reposent sur une organisation des images en mosaïque où l'utilisateur peut choisir, rejeter ou ne pas donner d'opinion, pour chacune des images présentées [HAL 05]. L'interactivité de cette navigation permet la formulation du besoin à partir de l'analyse d'images positives (choisies) et d'images négatives (rejetées). D'autres interfaces comme ImageGrouper [NAK 03] proposent aussi l'emploi d'images positives et d'images négatives. L'utilisateur navigue dans une collection d'images proposée aléatoirement ou à partir d'une image exemple et construit progressivement des groupes d'images qui peuvent être considérées comme positives ou négatives, ou neutres. Ces groupes peuvent être annotés, sauvegardés dans des albums et réutilisés pour une nouvelle recherche. Ces albums servent comme archives de résultats de requêtes. Ils peuvent être utilisés par d'autres utilisateurs pour amorcer une recherche.

Une autre forme de navigation, utilisant les graphes comme dans « TouchGraph Google Browser[3] », apparait pertinente pour une navigation référentielle par l'image. Dans ce cas, les nœuds sont des images et un arc reliant deux images matérialise une similarité entre ces deux images. Cette similarité peut être le résultat du calcul de similarité issu du modèle vectoriel (*Cf.* paragraphe 2.2) ou le résultat d'une autre forme de similarité basé soit sur tout ou une partie de l'indexation ou sur le contenu graphique des images.

Le graphe ainsi obtenu permet à l'utilisateur de naviguer dans les références imagées en sélectionnant un nœud et en déployant les nœuds (images similaires) de celui-ci.

La figure 2 propose une synthèse visuelle de ces formes de navigation et met en évidence les interactions ou cheminements possibles entre ces différentes formes.

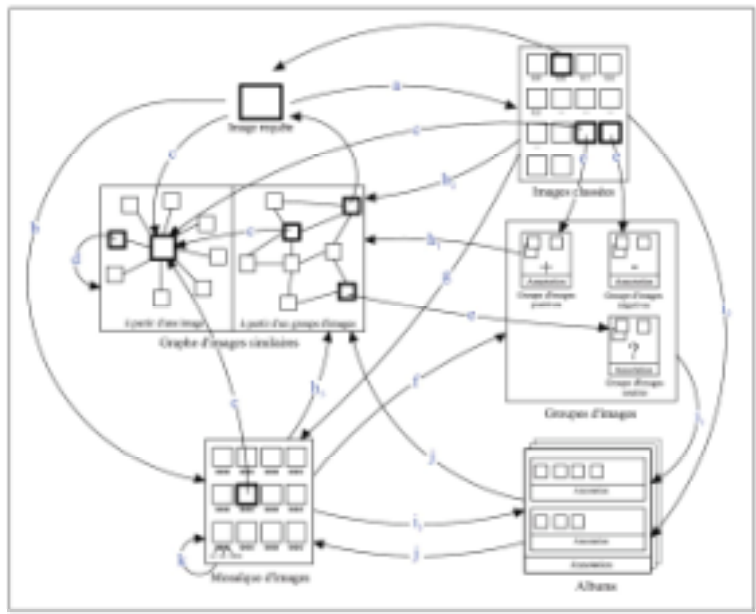

**Figure 2**. *Synthèse des modes de navigation référentielle par l'image*

– a : passage d'une image requête à un ensemble d'images classées par ordre de similarité ;

---

[3]http://www.touchgraph.com/TGGoogleBrowser.html.

– b : création d'une mosaïque d'image à partir d'une image requête ;

– c : organisation/visualisation des images similaires à une image requête sous forme d'un graphe d'images similaires ;

– d : déploiement des images similaires à une image sélectionnée dans un graphe ;

– e : création des groupes d'images positives, négatives ou neutres à partir d'un graphe d'images similaires ou à partir d'images classées ;

– f : passage d'un choix d'images dans une mosaïque d'images à des groupes d'images (respectivement) positives, négatives et neutres ;

– g : génération d'une mosaïque d'images à partir d'images classées ;

– h : visualisation de graphes d'images similaires à partir des groupes d'images ($h_1$), à partir d'images classées ($h_2$), à partir de la mosaïque d'image ($h_3$) ;

– i : construction d'albums d'images à partir des groupes d'images ($i_1$), à partir d'images classées ($i_2$), à partir de mosaïque d'image ($i_3$) ;

– j : amorce d'une recherche à partir d'un album d'images ;

– k : proposition d'une nouvelle mosaïque d'images à partir d'une première mosaïque où l'utilisateur a effectué des choix.

**4 Conclusion**

L'enjeu final de notre approche est de construire un outil de navigation référentielle par l'image pour assister la conception des ambiances lumineuses durant les phases initiales de la création. Dans cet article, nous avons exposé la méthode d'indexation d'images suivie. Cette méthode est basée sur une indexation sémantique. Chaque image est indexée par un ensemble de termes pondérés. La pondération permet de représenter les images sous une représentation vectorielle afin d'établir des similarités sémantiques entre les images.
Nous avons aussi présenté une analyse synthétique de différents modes de navigation par l'image référence. L'étape suivante est d'expérimenter ces modes de navigation sur la base de références que nous avons construite afin de proposer un outil d'assistance mettant en œuvre tout ou une partie de ces modes de navigation.

**5 Références**

[AMA 01] AMALIA T., « Indexation sémantique pour les systèmes de recherche d'informations », *Thèse de doctorat*, Université de Strasbourg 1, Strasbourg, 2001.

[CIR 91] CIRIANI H., « Lumière de l'espace », *Architecture d'Aujourd'hui*, no 274, 1991, p. 75-152.

[DAU 94] DAUZATS M., *Le thésaurus de l'image : étude des langages documentaires pour l'audiovisuel*, Paris, ADBS éditions, 1994.

[GOL 95] GOLDSCHMIDT G., *Visual display for design : Imagery, analogy and databases of visual images in Visual Databases in Architecture,* Aldershot, Avebury, Koutamanis A., Timmermans H., Vermeulen I., 1995.

[GRO 02] GROOT E., BERNARD P., « DIAL-Europe: New Functionality's for an Integrated Daylighting Design Tool », *6th Int. Conf. on Design Support Systems in Architecture and Urbanism Planning*, Landgoed Avegoor, juillet, 2002.

[GUI 92] GUIBERT D., *Du jeu des références,* Paris, PCA, 1992.

[HAL 05] HALIN G., BIGNON J.-C., HUMBERT P., KACHER S., « A method for Constructing a Reference Image Database to Assist with Design Process. application to the Wooden Architecture Domain », *International Journal of Architectural Computing*, vol. 3 no. 2, 2005, p. 227-243.

[KAC 02] KACHER S., HALIN G., BIGNON J.-C., DUFFING G., « The content based-image retrieval as an assistance tool to the architectural design domain », *Design and Decision Support Systems in Architecture and Urban Planning* Ellecom, The Netherlands, 7-10 juillet, 2002.

[KOU 07] KOUTAMANIS A., Halin G., Kvan T., « Indexing and retrieval of visual design representations », *eCAADe*, Francfort, 26-29 septembre, 2007.

[LAS 98] LASSANCE G., « Les configurations référentielles, un instrument conceptuel du projet d'ambiance », *Les cahier de la recherche architecturale, n°42/43* 3ème trimestre, 1998, Editions Parenthèses, p. 37-47.

[NAK 03] NAKAZATO M., MANOLA L., HUANG T. S., « ImageGrouper: a group oriented user interface for content-based image retrieval and digital image arrangement », *Journal of Visual Languages & Computing*, vol. 14 (4), 2003, p. 363-386.

[SAL 83] SALTON G., MC GILL M. J., *Introduction to Modern Information Retrieval,* International company New York, Mc Graw-Hill, 1983.

[VEL 01] VELTKAMP R. C., TANASE M., « Content-Based Image Retrieval Systems: A Survey », Utrecht University, 2001.

[VEN 06] VENTRESQUE A., « Recherche d'Information efficace utilisant la sémantique: le focus », *CORIA'2006*, Lyon, 15-17 mars, 2006, p 377-382.